\documentstyle[prc,aps,manuscript]{revtex}

\begin{document}

\title{ Multifragmentation of charge asymmetric nuclear systems. } 

\author{ A.B. Larionov$^{1,2}$, A.S. Botvina$^{3,4}$, M. Colonna$^5$,
M. Di Toro$^5$}

\address{$^1$ Institut f\"ur Theoretische Physik, Universit\"at Giessen,
         D-35392 Giessen, Germany}
\address{$^2$ Russian Research Center "I.V. Kurchatov Institute", 
         123182 Moscow, Russia}
\address{$^3$ INFN and Dipartimento di Fisica, Bologna, Italy} 
\address{$^4$ Institute for Nuclear Research, Russian Academy of Science,
         117312 Moscow, Russia}
\address{$^5$ Laboratori Nazionali del Sud, Via S. Sofia 44,
         I-95123 Catania, Italy\\ and University of Catania}

\maketitle

\begin{abstract}
The multifragmentation of excited  spherical nuclear sources with 
various $N/Z$ ratios and fixed mass number is studied within 
dynamical and statistical models. The dynamical model treats the 
multifragmentation process as a final stage of the 
growth of density fluctuations in unstable expanding nuclear 
matter. The statistical model makes a choice of the final multifragment 
configuration according to its statistical weight at a global thermal 
equilibrium. Similarities and differences in the predictions of the 
two models on the isotopic composition of the produced fragments are 
presented and the most sensitive observable characteristics are 
discussed.
\end{abstract}

\vspace{1cm}

\hspace{-\parindent}PACS numbers: 
25.70.Pq, 21.65.+f, 24.10.Cn, 24.10.Pa, 24.60.Ky

\newpage

\section {Introduction}

In recent years two main models, statistical and dynamical, have
been suggested to describe nuclear multifragmentation, i.e. the 
production of several intermediate mass fragments in heavy ion 
collisions. In the statistical picture a global equilibrium
is assumed in a freeze out volume and the various break-up channels
are selected in a microcanonical way according to the statistical weight
of the corresponding partition, given by the channel entropy 
\cite{Mis98,Gro97}. In the dynamical model clusters are directly
formed during the expansion phase by a collective response of the
interacting nuclear matter in low density unstable regions,
the spinodal decomposition mechanism \cite{Col98,CDG94}.

The standard applicability condition of a statistical approach
in nuclear reactions is a dominance of the available phase space
over transition matrix elements. Usually this will require 
a lot of interactions between nucleons and a related 
characteristic time scale which can be considerably
larger than the time of the growth of mean field instabilities
in the dynamical approach.  
Therefore one would
expect a corresponding validity for quite different physical scenarios
of formation of the highly excited source which will fragment
[see discussion in Ref.\cite{Col98}
and in the conclusions of Ref.\cite{CDG94}]. In fact the two models
seem to lead to very similar predictions for charge and kinetic energy
distributions of the same class of fragmentation events
\cite{Dur98,Riv98,Bou99}. One of the aims of this paper is to
suggest new observables in a study of isospin effects that
should help to disentangle between the two pictures.

Moreover the effect of charge asymmetry on fragment production is 
of large theoretical \cite{BBP71,Kon94,MS95,Liko97,RSK97,CDL98,BCDL98,Col99}
and experimental \cite{Yen94,Demp96,Kun96,KBK97,Sob97,Yen98,Mil99} 
interest. One important reason is that it could be 
related to the fundamental problem of extracting information on 
the nuclear equation 
of state (EOS), in particular on the symmetry energy, at subnuclear 
densities.

In this work, on the basis of both statistical and dynamical models,
we study the multifragmentation process in highly excited heavy
nuclei. Such systems can be created in central collisions of two
heavy ions at intermediate energies ($E_{lab} \sim 100$ MeV/nucleon) 
or in reactions induced by high energy ($E_{lab} \sim 1 $ GeV/nucleon) 
light projectile on heavy target. 
We present various measurable characteristics: charge yields, 
intermediate mass fragment $Z \geq 3$ (IMF) multiplicity distributions,
$N/Z$ ratio versus charge of fragment, isotopic ratios
in order to understand their sensitivity to the neutron excess of the 
source nucleus.

The interplay of symmetry and Coulomb energies defines the $N/Z$
ratio of formed fragments. Here and below, the {\it hot} fragments
produced just after multifragment breakup are discussed, if the 
opposite is not indicated. 
In the fast spinodal decomposition process
\cite{CDL98,BCDL98}, a higher density (liquid) phase becomes 
isotopically more
symmetric and a lower density (gaseous) phase accumulates the most part
of the neutron excess of the system. This effect is related
to the increase of the potential symmetry energy per nucleon
with the density at subnuclear densities $\rho \leq \rho_{eq}$,
$\rho_{eq}=0.17$ fm$^{-3}$, as given by all realistic effective
interactions \cite{Bomb,Prakash}.  
Therefore when the initially space-uniform low-density 
neutron-rich nuclear matter 
will clusterize the protons will be concentrated inside clusters. 
The fragment $N/Z$ ratio should decrease in this case with 
the charge number $Z$, since heavy clusters are associated with the liquid 
phase while light ones are created in a coalescence process taking place 
in the gaseous phase.
On the other hand
the Coulomb interaction, not present in the nuclear matter analyses
of ref.s \cite{CDL98,BCDL98} will act in the opposite direction 
preventing the proton clustering. Therefore it is important to work out
a simulation for a realistic finite nuclear source. We will use
a Stochastic Mean Field (SMF) approach \cite{Col98}, extension
of transport nuclear models to a consistent treatment
of fluctuations.
We have to stress that fragments produced in 
the {\it fast} spinodal decay will be sensitive to the density dependence of 
the symmetry energy.

In the case of finite nuclei in the stability valley we have in fact
a growth of the $N/Z$ ratio with the charge number $Z$.
One expects this behaviour of $N/Z$ also for fragments which are produced 
in the {\it slow} statistical process at thermal equilibrium,
since only the 
free energy of the system is important here, that is the sum of free energies
of fragments plus the energy of the Coulomb interaction between fragments.
In this case the memory of the dynamical features associated with the
spinodal decay is lost, leading to a chaotic evolution, where all possible
multifragmentation channels will be open \cite{Andrea}.
In the paper we will use the Statistical Multifragmentation Model (SMM)
(see Ref. \cite{BBIMS95}).   

Another important discrepancy in the predictions
of the two models is expected at the level of variances with respect to
mean values. Indeed we would have a
 much narrower isotopic mass distributions in the dynamical
description than in the statistical one, since the
strong thermal fluctuations, inherent
in the statistical approach where a wide range of channels can contribute,
will not be present in the dynamical model. 

These two striking discrepancies, 
on the background of an overall agreement in the predictions of the models
on other isospin-averaged observables, will be discussed below
on the basis of numerical results. 

\section{Comparison procedure: choice of the excited source}

It is known from earlier studies \cite{Jung88}, that in the mean field 
description a nucleus is very stable with respect to the multifragment 
breakup. In recent years stochastic extensions of mean-field
approaches have been introduced, well suited to describe
the so-called spinodal decomposition, i.e. a multifragmentation
process due to the occurrence of volume instabilities
\cite{CCR94,CC94}. In this picture the fragment
production is due to the growth of the most important collective unstable 
modes. In violent reactions at intermediate energy it is possible to 
observe the formation of a hot and
dense nuclear source, that subsequentely expands and enters the unstable
region of the nuclear matter phase diagram, the spinodal region. 
Within this scenario, because of the instability, the fluctuations of the 
local density are amplified, leading to a break-up of the nuclear system 
into many fragments. If the time scales of the expansion and of the 
instability growth are matching we will expect a fragmentation pattern 
with memory of the collective response of the unstable nuclear matter.
>From the above discussion it is clear that in a dynamical picture
part of the initial available energy should be put in a collective
radial flow.

In the following we will consider stochastic mean field (SMF) calculations.
Fluctuations are introduced at the initial time by agitating
randomly the density profile. The amplitude of these fluctuations is calculated
according to a Boltzmann-Langevin theory and depends on local density 
and temperature of the source \cite{Col98}.  

We have performed SMF calculations (see Ref. \cite{Col98}) 
of the multifragment break-up of $A=197$ isobars with $Z=95$, 
79 and 63. The temperature of the source nuclei was chosen to be
$T=3$ MeV, that is typical for multifragmentation in heavy ion
collisions. A radial flow energy of 3 MeV/nucleon was added in order
to prevent the resilience of the initially expanded source
($\rho_{init}^{SMF} = 0.5 \rho_{eq}$) to the normal density. 
The total excitation energy per nucleon, which
includes thermal, compressional and flow contributions was 
$E^*/A = 8$ MeV. The time evolution of the expanding source was followed
until 150 fm/c. At this time, the fragment formation process
due to the density fluctuation growth is already finished, and one can
extract various characteristics of hot fragments from the phase space
distribution, using a coalescence procedure, see Refs.\cite{Col98,Col92}.

We have performed 100 runs for each source nucleus. 
The accumulated statistics is enough to
calculate the charge yields of fragments, the average $N/Z$ ratios
and their fluctuations as a function of $Z$ (see Figs. 3,6,7). 

In parallel, we have applied the SMM model \cite{BBIMS95} to the same
sources taking also an overall excitation energy $E^*/A = 8$ MeV and 
a freeze-out volume 
$V_{f.o.} = 3 V_0$, where $V_0 = 4 \pi R_0^3 / 3$ is the volume of a 
ground state nucleus, $R_0 = 1.2 A^{1/3}$. 
We remark that the corresponding freeze-out
density $\rho_{f.o.}^{SMM} = \rho_{eq}/3$ is less than the initial density
$\rho_{init}^{SMF}$. 
However, in SMF the initial source expands while fragments are forming, so when
they are well defined (i.e. they do not feel anymore the nuclear interaction),
the volume inside which the intermediate mass fragments are located is larger
than the initial volume of the unstable source.
 
In Fig. 1 we show the radial distribution of the SMF-fragments at 
t=150 fm/c. The spatial ordering of fragments vs. their masses  
is clearly seen  at $r=7\div13$ fm, showing a bubble-like
distribution with larger fragments placed 
closer to the center. Notice, that the radius of the sphere containing 
the farthest fragments ($\sim 10$ fm) gives just nearly the freeze-out volume 
chosen for the SMM calculation. In the SMM the hot fragments are distributed 
more uniformly in the freeze-out volume, though a similar ordering trend is
observed because of conditions of finiteness and nonoverlapping of fragments.
The different space distributions of fragments should be also 
reflected in a different pattern of the Coulomb acceleration,
with on average a larger Coulomb interaction energy for the SMM fragments. 

In Fig. 2 we present the kinetic energy of fragments as a function of 
$Z$ for the SMF at t = 150 fm/c (dashed histogram) and after the
Coulomb acceleration (full histogram), and  for the SMM (stars -- 
before Coulomb acceleration, squares -- after Coulomb acceleration). 
The kinetic energy of heavy SMF-fragments is close to the one of the
SMM-fragments before Coulomb acceleration ($\sim 10$ MeV), which is
practically independent of $Z$ since thermalization is assumed in 
the statistical model. The kinetic energy of the SMF-fragments increases 
with $Z$ (for $Z\leq10$), reaches a maximum and then decreases at larger
$Z$.  This trend can be due to the Coulomb acceleration (that in the SMF 
calculations the fragments start to feel already before t = 150 fm/c) and
to the presence of a radial flow, though much reduced when compared to the
initial one. 
The decrease at larger $Z$ is due to the
preferential formation of heavy fragments closer to the center of
the system (see Fig. 1). 
The Coulomb acceleration of the SMF-configuration 
increases the kinetic energies of the SMF-fragments at the final time. 
However the final energies are 
below the SMM curve after Coulomb acceleration, due to the smaller Coulomb 
energy stored in the bubble-like SMF-configurations.

As it was pointed out in \cite{BBMS94} the statistical approach can
be used for the description of the fragment formation process even
at high energies once the preequilibrium emission leading to some 
mass and energy loss is taken into account. We have a similar effect
in the present SMF calculations: a part of nucleons escapes from
the fragment formation region fast and does not participate in 
the process. However this part is small in our case and does not
influence the final conclusions. 
  
In Fig. 3 the fragment charge yields are presented for both models.
We see that slopes and absolute values of the charge yields for 
statistical and dynamical models are very similar for $Z > 5$. 
This is indeed a good trigger for a meaningful comparison
of the fragment properties.
We remark that in the spinodal decomposition mechanism the 
direct production of primary light clusters is reduced due to the 
finite range of the nuclear force which prevents the
the propagation of short unstable wave lengths \cite{Col98}.
Moreover the study of light clusters is out of the scope
of mean-field approaches, so in the following we will
concentrate on IMF ($Z \geq 3$) properties.
The results for {\it cold} fragments (Fig.3d) have been obtained
with an evaporation decay described in the next section.

The IMF yield is roughly proportional to the charge number of the 
fragmenting source. This can be seen from the IMF multiplicity 
distributions presented in Fig. 4. The effect is weaker in the
SMF results. An explanation could be 
a very weak dependence of the wavelength of the most unstable mode on
the charge asymmetry of the system (see Ref. \cite{CDL98,BCDL98}). 
In general the SMM predictions give larger yields 
for primary IMFs, mostly due to the enhanced yield of fragments
with $Z \leq 5$ discussed above (see Fig. 3). We remind
that the IMF has a charge number $Z \geq 3$ by its definition.

\section{Fragment properties and isotopic ratios}

In order to compare the energy balance in dynamical and statistical
calculations we present in Fig. 5 the excitation energies of the
hot fragments. For instance, we see from Fig. 5c, that for the initial 
source with larger asymmetry, $Z=63$, the primary SMM-fragments are 
excited stronger than the SMF-ones. This is interesting since we would 
expect the opposite after the discussion on kinetic energies presented 
in Fig. 2, since SMM-fragments have also larger kinetic energies. 
This is a consequence of two main
effects: (i) In the dynamical evolution a large fraction of energy
is dissipated in the neutron-rich gas and the primary fragments of
the liquid phase are less excited; (ii) In the statistical
picture an increasing share of surface contributions
in light IMFs causes a slight increase 
of their excitation energy per nucleon.
However in general the difference between the fragment excitation 
energies from SMF and SMM calculations is always no more than 
0.5 MeV/nucleon. 

\subsection{Mean values}

In Fig. 6 the average ratio $\langle N/Z \rangle$  of produced hot 
(a,b,c) and cold (d) fragments is shown versus the fragment charge number 
$Z$. For the almost symmetric source with $Z=95$, (a), fragments have 
practically the same initial $N/Z$ (see arrow). 
For the gold fragmentation $Z=79$, (b), the fragments have slightly lower 
$N/Z$ ratio with respect to the one of the initial source due to the 
enhanced neutron emission during the multifragment breakup.
This reduction of the $N/Z$ of hot fragments becomes more pronounced for
the mostly neutron rich source $Z=63$, (c). In this case we observe also
a clear effect of the density dependence of the symmetry energy:
dynamical calculation (solid histogram) gives the decrease of the $N/Z$ 
with charge number $Z$ at $3 \leq Z \leq 10$, while the statistical 
calculation shows the increase of this ratio for primary fragments,
as it was discussed above in the Introduction. Unfortunately, this effect 
seems to be completely washed out in the $\langle N/Z \rangle$ ratio versus 
$Z$ dependence for {\it cold} fragments, Fig.6d.

The same secondary deexcitation procedure implemented in SMM 
(see Ref. \cite{BIMB87}) 
was applied for the cooling of the hot fragments produced either 
dynamically or statistically. In the final products there is an universal 
decrease of the $\langle N/Z \rangle$ with $Z$, due
to a larger Coulomb barrier for the emission of light charged particles
by heavier hot fragments. That results in the increased relative 
contribution of the neutron emission with respect to the charge particle
emission for the deexcitation of heavier hot fragments. The fact that all
calculations produce the same dependence $\langle N/Z \rangle(Z)$ for cold 
fragments leads to the conclusion that the excess neutrons of hot 
fragments present in the SMF and SMM results with initial source $Z=63$
(Fig. 6c, solid histogram and diamonds) are very weakly bound.
They can then escape
from the compound nucleus at the very first steps of the 
deexcitation process.
This first-chance neutron emission almost does not change the excitation
energy of the compound system. Therefore, the subsequent de-excitation steps
proceed almost in the same way in all calculations. 
Of course in the evaporation chains
we have now to insert nuclear structure properties of very exotic
nuclear systems, which are largely unknown. So we can expect a quite
large uncertainty in the predictions of the secondary deexcitation
described using the method of Ref.\cite{BIMB87} or any other
sequential decay codes. Therefore we believe that the performed 
calculations can provide some guidelines for an experimental investigation
of this phenomenon.

\subsection{Dispersions}

The difference between dynamical and statistical calculations is
particularly pronounced in the dispersions of the $N/Z$ ratio for
each $Z$ (Fig. 7). For hot fragments, the statistical calculations 
produce a factor of $2\div5$ larger dispersion even in the cases when 
there is no difference in the mean values of the $N/Z$ (compare Fig. 6 
a,b to Fig. 7 a,b). 
As we will see in the following, the variances are so large that 
the results presented below are mostly sensitive to that and not 
much depending on 
the mean values. 
Physically, this difference is related to a more 
deterministic way of the fragment formation in the SMF, that prohibits the 
strong deviations of the $N/Z$ ratio from its mean value. Again, the 
subsequent deexcitation of the hot fragments seems to 
wash out completely all the differences in the cold IMFs (see Fig. 7d).

The previous discussion was mostly concentrated on the properties
of hot primary fragments. We think it is possible to find some experimental 
observables which are sensitive to the differences between dynamical 
and statistical results for the hot fragments by looking at the products 
of their secondary decay. There is already an existing experimental
technique based on a correlation function analysis that allows to
attribute a secondary light charged particle to an IMF from which it was 
emitted \cite{Marie98}. 

In Fig. 8 we plot the yield ratios of $^4He$ 
to $^3He$, (a), and $^3H$ to $^3He$, (b), 
only emitted from the primary IMFs,
as a function of the $N/Z$ of the source. In this way we avoid the 
problem
of preequilibrium particles and particles directly produced in the 
multifragment breakup. The dynamical calculation
gives always larger ratios than the statistical one.  The difference 
increases for the high asymmetric source 
($A=197$, $Z=63$, $N/Z=2.13$). This difference is due to a larger
dispersion of the $N/Z$ ratio for the hot SMM fragments.  
In the neutron excess region the production of a proton rich light 
particle becomes extremely sensitive to small changes of the $N/Z$ of 
the primary fragments. Therefore, larger fluctuations of the $N/Z$ ratio 
of hot fragments in the statistical calculation (see Fig. 7c) cause a 
dramatic enhancement of the $^3He$ production with respect to the dynamical 
calculation.

\section{Conclusions}

We have performed a comparative study of the multifragmentation of 
highly excited asymmetric spherical sources within the dynamical and the 
statistical models. In spite of the general coincidence of the results, 
two clear differences in the isotopic composition of the hot primary 
fragments for a neutron rich initial source
have been observed: (i) in the dynamical calculation the average $N/Z$ 
ratio decreases with charge number $Z \geq 3$, while in the statistical
one it increases; (ii) the dispersion of the $N/Z$ ratio for a given 
charge number $Z$ of hot fragment is much larger in the statistical 
calculation than in the dynamical one.
Both effects seem to be largely reduced in the final fragments
seen in the detectors. There is however a chance that some
signatures of differences in isotopic content and excitation energy
of primary fragments could survive. A good observable seems to be
the isotopic ratio of light charged particles emitted from hot
primary fragments.

We remark that, as already stressed in the Introduction, the point
i), i.e. more symmetric primary fragments produced with the
spinodal mechanism, is present for all effective interactions:
the {\it magnitude} of the effect is sensitive to different 
behaviours of the symmetry term at {\it subnuclear} densities
\cite{CDL98}. 

To disentangle between the 
two models a comparison of fragmentation data for
some extremely neutron rich sources $N/Z \sim 2$ is required.
A good possibility would be to look at the fragment produced
in the midrapidity region in semiperipheral collisions of
neutron excess nuclei: taking advance of the "neutron skin" effect
we could study properties of fragments produced from a highly
asymmetric interacting nuclear matter.

Some recent data on charge asymmetry effects on fragment production
seem to be in a qualitative agreement with the signatures of the
dynamical model, although a more precise analysis must be performed:

1) More neutron rich systems are producing more symmetric and less
excited primary intermediate mass fragments \cite{Kun96,KBK97,Mil99};

2) Such effects are disappearing for central collisions above $100~AMeV$
beam energy \cite{Mil99}: in the dynamical picture the expansion phase
becomes too fast and the instability growing times cannot be matched;

3) Opposite behaviour is observed for light ions produced in the
midrapidity region, that appear more neutron rich than the average
$N/Z$ of the composite system \cite{Demp96}.

It should be mentioned here that the result 1) can be obtained also in 
recently developed thermodynamical models \cite{MS95,Frafil}.

If all these signatures will be confirmed,
from medium energy heavy ion collisions we will have a new
powerful method to study the symmetry term of the nuclear
Equation of State at subnuclear densities, of large
interest for the structure of unstable nuclei as well as
for astrophysics \cite{PR93} 

\section*{ Acknowledgements }

We are grateful to U. Mosel, W. Cassing and H. Lenske for fruitful 
discussions. 
Two of us (A.B.L.) and (A.S.B.) acknowledge the hospitality and support 
of INFN (LNS-Catania and Bologna) where the most part of this work was 
done, and one of us (A.B.L.) acknowledges the support of BMBF and GSI 
and the hospitality of the Institut f\"ur Theoretische Physik at the
University of Giessen.

\newpage

\section*{ Figure captions }

\begin{description}

\item[Fig. 1] Number density of hot fragments as a function of distance 
from the center of source with $Z=63$ for the SMF calculation at the time 
step $t=150$ fm/c. Solid, short- and long-dashed lines correspond to
fragments with charge numbers $Z=1\div2$, $3\div10$ and $\geq11$ 
respectively.   

\item[Fig. 2] Kinetic energy of a fragment as a function of the charge 
number for the source with $Z=63$. Dashed (solid) histogram and stars
(squares) show SMF and SMM results before (after) Coulomb acceleration. 

\item[Fig. 3] Charge yields of hot (a,b,c) and cold (d) fragments normalized 
on the number of fragments per event. Histogram and diamonds -- SMF and SMM 
results respectively with source $Z=95$ (a), 79 (b) , 63 (c,d). 

\item[Fig. 4] IMF multiplicity distributions. Notations of curves 
and panels a,b,c,d like in Fig. 3.

\item[Fig. 5] Mean excitation energy per nucleon versus fragment
proton number.  Notations of curves and panels a,b,c like in 
Fig. 3.

\item[Fig. 6] Mean neutron-to-proton ratio $\langle N/Z \rangle$ 
of fragments versus proton number $Z$. Notations of curves and 
panels a,b,c,d 
like in Fig. 3. Arrows show $N/Z$ ratios of initial sources with $Z=95$ 
(a), 79 (b) and 63 (c). 

\item[Fig. 7] Dispersion 
$\sigma_{N/Z} = \sqrt{ \langle ( N/Z - \langle N/Z \rangle )^2 \rangle }$ 
as a function of the fragment proton number $Z$.  Notations of curves and 
panels a,b,c,d like in Fig. 3.  

\item[Fig. 8] Isotopic ratios $^4He/^3He$ (a) and $^3H/^3He$ (b) 
are plotted as a function of the $N/Z$ of the source. Solid lines 
with errorbars show the SMF result (errors are statistical ones).
Diamonds represent the SMM calculation.
     
\end{description}

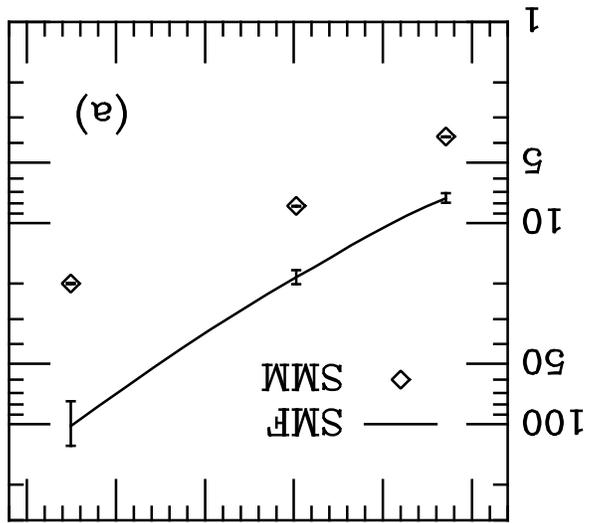
\end{document}